*Dedicated to the memory of Dr. Mikhail Eremets, our friend, who discovered superconductivity in hydrides and made a key contribution to their studies. One of the authors (VMP) had the honor to participate in the conference and ceremony dedicated to the awarding of M. Eremets with the Ugo Fano medal in 2015.*

**Stability and Superconductivity of Ternary Polyhydrides**

*Dmitrii V. Semenok*, Di Zhou*, Wuhao Chen, Alexander G. Kvashnin, Andrey V. Sadakov, Toni Helm, Pedro N. Ferreira, Christoph Heil, Vladimir M. Pudalov, Ivan A. Troyan, and Viktor V. Struzhkin**

Dmitrii V. Semenok, Di Zhou
Center for High Pressure Science & Technology Advanced Research, Bldg. #8E, ZPark, 10 Xibeiwang East Rd, Haidian District, Beijing, 100193, China
E-mail: dmitrii.semenok@hpstar.ac.cn, di.zhou@hpstar.ac.cn

Viktor V. Struzhkin
Shanghai Key Laboratory of Material Frontiers Research in Extreme Environments (MFree), Shanghai Advanced Research in Physical Sciences (SHARPS), 68 Huatuo Rd, Bldg #3 Pudong, Shanghai 201203, China
9 Center for High Pressure Science & Technology Advanced Research, 1690 Cailun Rd, Bldg #6, Pudong, Shanghai 201203, China
E-mail: viktor.struzhkin@hpstar.ac.cn

Wuhao Chen
Quantum Science Center of Guangdong–Hong Kong–Macao Greater Bay Area (Guangdong), Shenzhen 518000, China

Alexander G. Kvashnin
Skolkovo Institute of Science and Technology, 121205, Bolshoy Boulevard 30, bld. 1, Moscow, Russia

Andrey V. Sadakov, Vladimir M. Pudalov
V. L. Ginzburg Center for High-Temperature Superconductivity and Quantum Materials, 53 Leninsky Prospekt, building 10, Moscow 119991, Russia

Toni Helm
Hochfeld-Magnetlabor Dresden (HLD-EMFL) and Würzburg-Dresden Cluster of Excellence, Helmholtz-Zentrum Dresden-Rossendorf (HZDR), Bautzner Landstraße 400, Dresden 01328, Germany

Pedro N. Ferreira, Christoph Heil,
Institute of Theoretical and Computational Physics, Graz University of Technology, NAWI Graz, 8010, Graz, Austria

Ivan A. Troyan
A.V. Shubnikov Institute of Crystallography of the Kurchatov Complex of Crystallography and Photonics, 59 Leninsky Prospekt, Moscow 119333, Russia

Funding:
National Natural Science Foundation of China (NSFC), grant No. 12350410354.
China Postdoctoral Science Foundation, No. 2023M740204.
Shanghai Science and Technology Committee, China, No. 22JC1410300.
Shanghai Key Laboratory of Materials Frontier Research in Extreme Environments, China, No. 22dz2260800.
National Key Research and Development Program of China No. 2023YFA1608900, subproject 2023YFA160890.
Russian Science Foundation, No. 24-72-10109.

We review five years of experimental and theoretical attempts (2020–2025) to enhance the superconducting critical temperature ($T_c$) of hydrogen-rich compounds by alloying binary superhydrides with additional elements. Despite predictions of higher $T_c$ in ternary systems such as La–Y–H, La–Ce–H, and Ca–Mg–H, experiments consistently show that the maximum Tc in disordered ternary superhydrides does not exceed that of the best binary parent hydrides within experimental uncertainty. Instead, alloying primarily stabilizes high-symmetry polyhydride phases at lower pressures, enabling $T_c \approx 200$ K near 100–110 GPa, while also improving vortex pinning and upper critical fields. Magnetic dopants suppress $T_c$, whereas nonmagnetic additives leave it nearly unchanged, reminiscent of Anderson's theorem. These findings indicate that alloying is unlikely to raise $T_c$, but can reduce the pressures required to stabilize high-$T_c$ phases. We propose that fully ordered ternary hydrides, synthesized via controlled hydrogenation of intermetallic precursors, offer a promising route toward this goal. One of the most promising compounds of this kind is the recently discovered $LaSc_2H_{24}$.

## 1. Introduction

Conventional superconductivity is remarkably tolerant to disorder. In 1958–1959, Anderson, Abrikosov, and Gor'kov demonstrated that isotropic, nonmagnetic impurities leave both the s-wave superconducting gap Δ(0) and the critical temperature $T_c$ essentially unchanged, provided their concentration does not exceed a critical threshold [1-3]. The range of validity of this result for isotropic metals is defined by the condition

$$\frac{h}{\tau E_F} = \frac{n {m_e}^{3/2}}{2\pi^2\sqrt{2E_F}} \int |U(\bar{p}-\bar{p}')|^2 d\Omega_{p'} \ll 1, \qquad (1)$$

where τ – is the collision time resulting for the impurity potential in the Born approximation, $E_F$ – is the Fermi energy, $n$ – is the impurity concentration, $m_e$ – is the electron mass, $U$ – is the electron potential energy in the vicinity of the impurity atom in momentum space, and integration is performed over the Fermi surface [4]. Experimental evaluations of the collision time ($\tau \sim 10^{-15}$ s for about 10 at. % of Nd in $LaH_{10}$ [5]) show that at $E_F$ = 1–10 eV this condition of validity of Anderson's theorem (1) is fulfilled.

Superhydrides are probably well-suited for exploring this theorem because they demonstrate relatively weak pressure (and, hence, volume) dependence of the critical temperature, specifically in the pressure region of the maximum $T_c$, where the flat part of the dome-like $T_c(P)$ is observed [6]. These relatively flat regions of the $T_c(P)$ dependence were first discovered by M. Eremets' group for sulfur [7] and lanthanum [6,8] hydrides, and subsequently confirmed for yttrium[9,10], cerium[11], lanthanum-cerium[12,13] and other hydrides [14]. Notably, this dome-shaped dependence mirrors the characteristic variation of $T_c$ with doping observed in cuprate and iron-pnictide superconductors [15].

Despite the fact that the pressure and impurities change the size of the unit cell, even distorting its symmetry, the critical temperature in the vicinity of the highest $T_c$ value is usually stable to this within ± 5%. For instance, if $T_c$ = 200 K, then it varies only by $\Delta T_c \approx \pm$ 5 K, within the interval ± 15 GPa, that is close to the accuracy of experimental measurements of $T_c$ in DACs (see Table 4 in Ref. [16]). In conventional low-$T_c$ superconductors, even a 5 K change in $T_c$ is significant, since $T_c$ rarely surpasses 23 K (except for $MgB_2$). A 5–10 K difference, however, is within the typical experimental uncertainty for superconductivity in high-pressure hydrides, which matches with the remarkably consistent the critical temperature values reported by independent groups (see Table 4 in Ref. [16]).

Hydrides are synthesized from metals/lower hydrides precursors and are inevitably covered to various degree with oxide films. Furthermore, the use of $NH_3BH_3$ as a hydrogen source, with a high probability gives inclusions of B, and N in the compound, whereas the introduction of pure $H_2$ into the working volume between diamond anvils may cause carbon and $CH_4$ impurities from diamond anvils [17], as well as the presence of hydrogen vacancies.

In this paper we analyze the known experimental attempts to improve the properties of binary hydrides by introducing a third element into the system. As we demonstrate, for all known examples with a disordered sublattice, A-B-H systems yield the maximum critical temperature

$$\max T_c(\text{A-B-H}) \leq \max\,[T_c(\text{A-H}),\ T_c(\text{B-H})], \qquad (2)$$

not exceeding maximum of [$T_c$(A-H), $T_c$(B-H)] for binary systems within 5% experimental error. In our opinion, the main result of five years (2020-2025) of experimental research on ternary hydrides with disordered sublattices of heavy atoms (also called “pseudo ternary hydrides”) was the discovery of the effect of the so called “superhydride stabilization”. Impurities, increasing the entropy of compounds, stabilize some phase modifications (for example, the cubic high-temperature phases $XH_9$ and $XH_{10}$) at a lower pressure than it is observed for pure binary systems. Now we can expect to obtain in hydrides $T_c$ of about 200-230 K at a pressure of only 100-110 GPa, which significantly increases the range of experiments that can be conducted on such samples.

## 2. Results

### 2.1. La-Y-H system

La-Y-H ternary system was the very first investigated in 2020–2021, almost immediately after the discovery of superconductivity in $LaH_{10}$ and $YH_6$. The motivation was that *fcc* $YH_{10}$ had been predicted to be a room-temperature superconductor with a $T_c$ up to 326 K [18,19], which is significantly higher than in $LaH_{10}$ [6]. The problem, however, was that $YH_{10}$ turns out to be thermodynamically unstable and cannot be synthesized in the same way as *fcc* $LaH_{10}$ [9].

Given the similarity in chemical properties of La and Y, it was natural to expect that Y impurities in the $LaH_{10}$ lattice would substitute the La atoms and would be located in the correct positions of the hydrogen clathrate lattice, see Figure 1a. To shed light on the expected behavior of the (La,Y)$H_{10}$ solid solution, we present in Figure 1b,d preliminary ab initio results for $La_{1-x}Y_xH_{10}$ at 250 GPa, obtained within the Extended Generalized Quasichemical Approximation (EGQCA) recently introduced by Ferreira et al. [20] for superconducting alloys. In the EGQCA formalism, an alloy is represented as an ensemble of nonequivalent supercells (clusters) spanning the full compositional range. These supercells are treated as energetically and statistically independent of their local atomic environments, while the system remains macroscopically homogeneous. The occurrence probability of each cluster is determined by free-energy minimization, and any composition- and temperature-dependent property accessible from first principles is obtained as the ensemble average weighted by these cluster-occurrence probabilities.

We performed simple, preliminary EGQCA calculations for $La_{1-x}Y_xH_{10}$ at 250 GPa by computing all nonequivalent configurations within the 1×1×1 cubic unit cell. Thermodynamically, the system approaches the ideal random-cluster limit, as shown in Figure 1b by the Gibbs mixing free energy from 300 to 1400 K. No common-tangent construction connects distinct compositions in $\Delta G$, which rule out binodal and spinodal phase decompositions above ambient temperature. Thus, setting aside dynamical stability considerations for the moment, we would expect that homogeneous $(La,Y)H_{10}$ solid solutions at 250 GPa should be synthesizable with relative ease.

Figure 1d shows the evolution of $T_c$ with composition. To obtain $T_c$, we rescaled the Eliashberg spectral function $\alpha^2F(\omega)$ (Figure 1c) according to the density of states at the Fermi level computed with the tetrahedron method [21]. We solved the Migdal–Eliashberg equations [22] within the full-bandwidth approximation [23] as implemented in the IsoME code [24]. The Coulomb pseudopotential $\mu^*$ was renormalized to the Matsubara-frequency energy scale as described in Ref. [25]. Assuming a composition-independent $\mu^*$, $T_c$ of the $(La,Y)H_{10}$ increases nearly linearly up to $x \sim 0.75$, reaching 271 K for $YH_{10}$. This conclusion is consistent with other theoretical calculations for both $YH_{10}$ [18,19] and true ternary compounds predicted in the La–Y–H system [26]. However, if $\mu^*$ is taken to be larger in $YH_{10}$ and to increase linearly from La to Y, $T_c$ remains almost constant up to $x = 0.75$, and the $T_c$ of $YH_{10}$ is reduced to 242 K. Interestingly, accounting for the uncertainty in the theoretical estimate — quantified by the standard deviation of the weighted-ensemble distribution, which reflects thermal effects, and is shown by the blue and red shaded regions in the plot — reconciles theory and experiment.

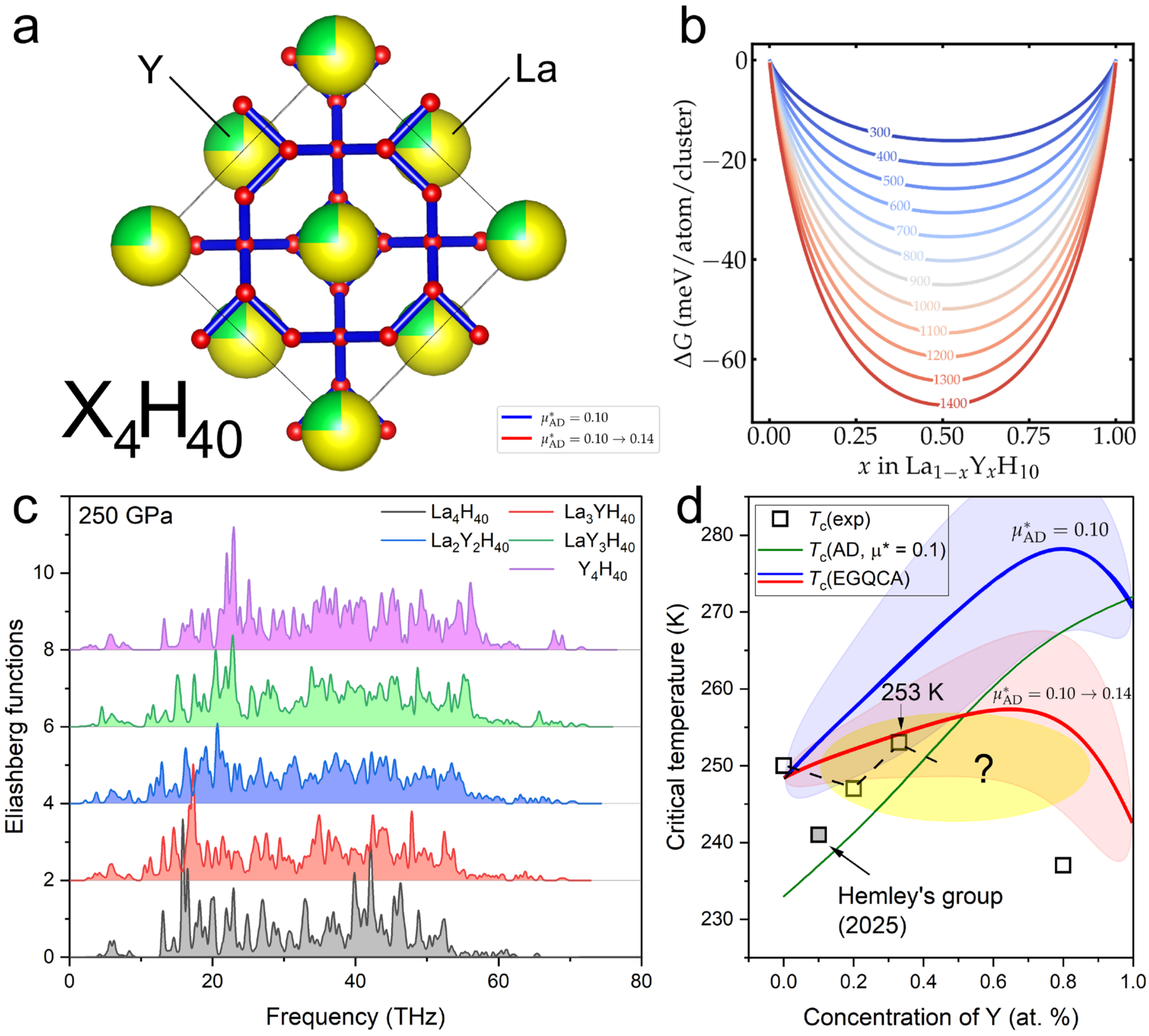

**Figure 1.** Theoretical and experimental investigation of superconductivity in (La,Y)$H_{10}$ superhydrides. (a) A schematic representation of the *fcc* $X_4H_{40}$ crystal structure, highlighting the positions of Lanthanum (La, yellow), Yttrium (Y, green), and Hydrogen (red) atoms. (b) Computed change in Gibbs free energy (ΔG) as a function of Yttrium concentration ($x$) in $La_{1-x}Y_xH_{10}$ for various temperatures. This includes also mixing entropy contribution. (c) Eliashberg functions for different compositions of (La,Y)$H_{10}$ at 250 GPa calculated using smearing method (σ = 0.05 Ry). (d) Superconducting critical temperature ($T_c$) as a function of Yttrium concentration in $La_{1-x}Y_xH_{10}$. An experimental data point from R. Hemley's group (2025) [27] showed separately. A region of uncertain behavior is also indicated (marked as "?"). The plot compares experimental data points (squares) with theoretical DFT predictions (lines) based on Eliashberg function calculations. Standard deviation of the weighted-ensemble distribution, which reflects thermal effects within the EGQCA approach, is shown by the blue and red shaded regions in the plot. The red curve corresponds to a linear change in μ* from 0.1 ($LaH_{10}$) to 0.14 ($YH_{10}$) and is shown only to demonstrate the possible effect of the dependence of $\mu^*(x)$ on $T_c$.

Surprisingly, in two independent experiments [26,27], no significant increase of $T_c$ with respect to $T_c$ of $LaH_{10}$ was observed over a wide range of yttrium concentration from 10 to 75 at. %. On the contrary, when the Y content in the La–Y alloy exceeded 33 at. %, the critical temperature of the corresponding hydride decreased from about 253 to 237 K and lower [26]. We found that the experimental trend of $T_c$ vs Y-concentration is unexpectedly well described within the framework of the Anderson's theorem. Of course, we also admit other explanations for the observed inability to improve the superconducting properties of hydride superconductors, for example, the trade-off between λ and $\omega_{log}$ upon doping or the behavior of $\mu^*$ as a function of composition. However, accurately capturing these effects requires accounting for anharmonicity, Coulomb screening, anisotropy, and other factors that are not fully included in most of DFT calculations of superhydrides.

Moreover, if the Y is replaced with the even smaller scandium atom (La–Sc–H system), then, as we recently showed, no changes in the superconducting critical temperature happens and the maximum well-reproducible $T_c$ value in $(La,Sc)H_{10-12}$ remains about 246 K [28].

There are several reasons for this discrepancy between the theory of La-Y-H system and the experiment:

- **Lattice dynamics.** Calculations within the harmonic approximation show that clusters with 25%, 50%, and 75% Y content in $La_{1-x}Y_xH_{10}$ are dynamically unstable, especially at pressures below 200 GPa. Such instabilities distort the hydrogen sublattice and break cubic symmetry, reducing $T_c$. Beyond this, anharmonic effects — which are not fully captured in most calculations — are expected to further lower $T_c$ in superhydrides.
- **Electronic screening.** The nature of Coulomb screening in these compounds, and its sensitivity to disorder, remains poorly understood. In particular, the effective Coulomb pseudopotential $\mu^*$ may vary with composition, and different assumptions about $\mu^*(x)$ lead to qualitatively different $T_c$ trends.
- **Sample effects.** Experimental samples inevitably contain hydrogen vacancies, oxide contamination, and phase coexistence due to laser heating or diffusion-limited growth. These imperfections can stabilize off-stoichiometric structures such as $XH_9$ and suppress $T_c$ relative to predictions for idealized ordered phases.

Thus, the La–Y–H system has been studied both experimentally and theoretically over a wide range of concentrations and pressures. There remains a potentially interesting region of Y concentrations between 50 and 70 at. %, which has not yet been investigated, and where at pressures of 200–250 GPa there is a chance of detecting higher $T_c$ value due to stabilization of $YH_{10}$.

At present, our group is exploring the creation of complex topologies on diamond substrates from active hydride-forming metals (La, Y, Ce, Th …) and their alloys. Such topologies, through subsequent laser heating or passive hydrogen diffusion, can be converted into superconducting patterns with high critical $T_c$. One of the simplest topologies is a ring-shaped sample (Figures 2a–c), the center of which, under favorable circumstances and laser-synthesis defects, may trap quantized magnetic flux like a superconducting quantum interference device (SQUID) [29].

The pathway to realizing such topologies includes deposition of thin films of active metals, followed by their transfer into a Focused Ion Beam (FIB) device, loading with ammonia borane, soldering of wires, etc. (Figure 2b, a ring with electrodes). An alternative process option is electron beam lithography. Here, the study of lanthanum alloys is particularly useful: La and La–Ce oxidize very easily in air, while La–Sc, La–Y, La–Th, La–Mg, La–Zr, etc. are much more stable and can be used both for sputtering and loading into high-pressure DACs without a glovebox.

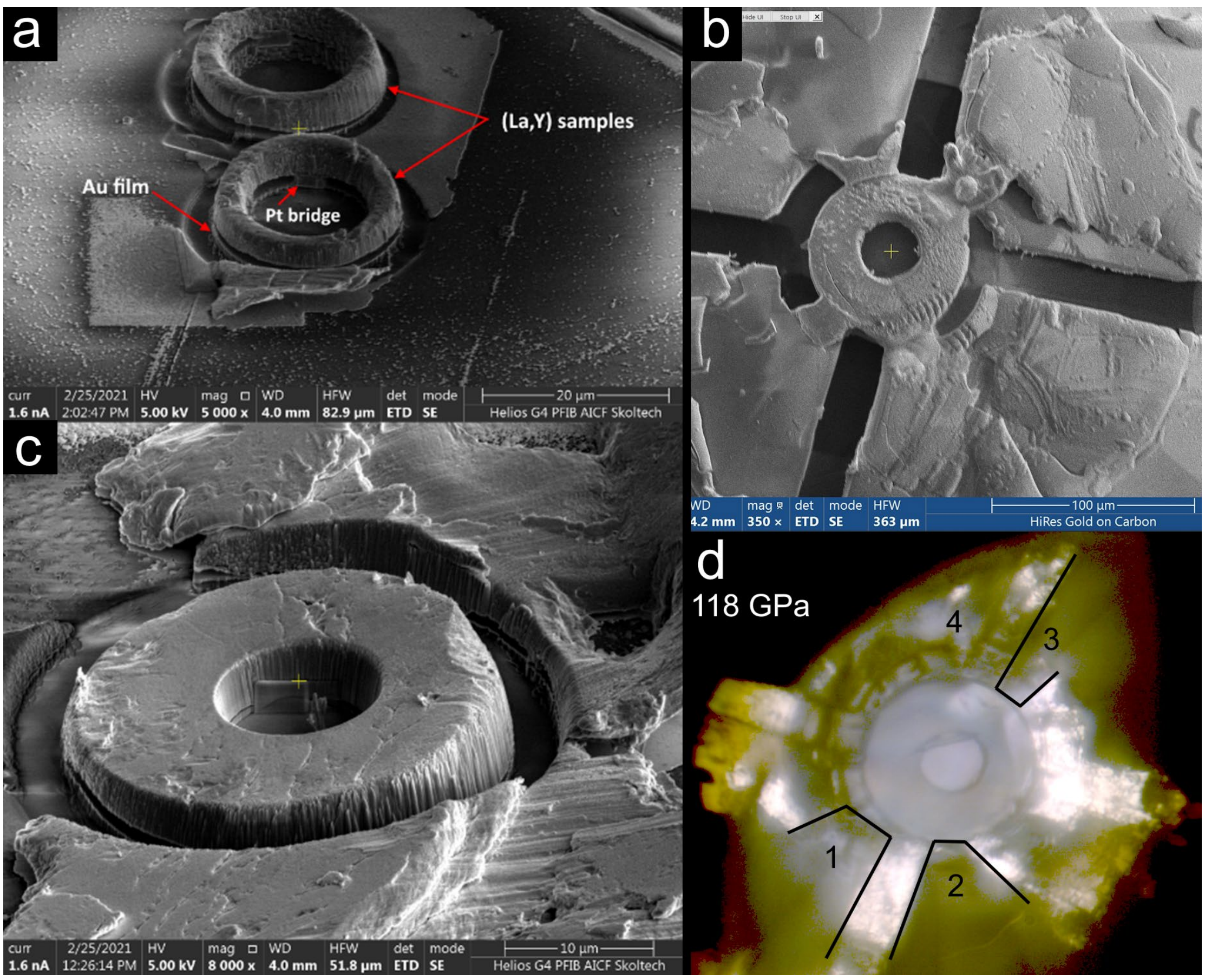

**Figure 2.** Focused Ion Beam (FIB) cutting of a (La,Y) alloy films for high-pressure electrical transport measurements in a ring geometry (for SQUID). (a) A scanning electron microscope

(SEM) image showing two ring-shaped (La,Y) samples connected by a platinum (Pt) bridge on top of a gold (Au) film, all fabricated using the Xe FIB technique. Yttrium protects Lanthanum from oxidation in air. (b) A high-magnification SEM image of a micromachined (La,Y) ring fixed by Pt soldering. (c) Another SEM image from a different perspective, highlighting the precise, three-dimensional cut of sputtered 1 μm film of (La,Y) alloy made by the Ga FIB. Ring is connected to four sputtered electrodes. (d) An optical image of the final sample assembly in a DAC's chamber at 118 GPa, showing the four electrical contacts (labeled 1-4) for resistance measurements. It is clearly visible that the sprayed ring can withstand compression above 1 Mbar.

**2.2. La-Ce-H system**

The La–Ce–H system has been studied quite thoroughly in four experimental works [12,13,29,30]. This system attracted attention after the discovery of the unique stability of $CeH_9$, which remains a high-temperature superconductor with $T_c \approx 80$ K even below 100 GPa [11] and exhibits a high critical current density. For example, applying a current of 100 mA reduces $T_c(CeH_9)$ by only 5 K and is accompanied by only minor broadening of the superconducting transition (Figure 3a). The pronounced and narrow superconducting transition detected at such a high current confirms the bulk nature of superconductivity in the $CeH_9$ without the need for any magnetic measurements.

Adding cerium to lanthanum increases the stability of the ternary hydride derived from them $(La,Ce)H_{9-10}$ [12,13]. However, at the same time, it leads to strong suppression of superconductivity from 250 K to around 200 K (Figure 3b) due to spin-flip scattering in the interaction of Cooper pairs with the local magnetic moment of Ce atoms, which have *f*-electrons in the outer shell. The level of $T_c$ suppression can be estimated from the derivative $dT_c/dx = -1.24$ K/$Ce_{\%}$ (Figure 3d).

Nevertheless, by adjusting the Ce concentration in the range of 10–25%, one can retain a superconducting $T_c$ near 200 K at pressures of 110–130 GPa (see Ref. [29] and Figure 3b), which makes it possible to use standard 100 μm diamond anvils well below their upper pressure limits. This nearly doubles the $T_c$ compared with pure $CeH_9$–$CeH_{10}$ (Figure 3a), enables to increases the sample's volume, its area, and improves the success rate of experiments by lowering synthesis pressure. Recent progress in observing advanced effects such as asymmetric conductivity and SQUID-like behavior in $(La,Ce)H_{10+x}$ is directly related to the relatively large sample sizes, in which many inhomogeneities and even holes appear after laser heating with hydrogen depletion [29].

Moreover, recent experiments with (La,Ce) hydrides using contactless radio-frequency detection of $T_c$ [30] showed that $T_c$ in La–Ce hydrides can reach 220–240 K (Figure 3c) and temporarily even rise higher. Certainly, $T_c$ > 250 K can hardly be attributed to the $XH_{10}$ phase, but the formation of a high-temperature superconducting $XH_{12}$ phase, for example, $CeH_{12}$ or $LaH_{12}$, cannot be excluded. Such phases were previously found in the La-H [6], Ba-H [31], La-Sc-H [28] and predicted for several other elements ($LuH_{12}$ [32], $ScH_{12}$ and $MgH_{12}$ [33]).

As already mentioned, Ce-atoms have $d^1f^1$ electron configuration and introducing such magnetic impurities into the lattice quickly suppresses $T_c$ (see Fig. 3d, $dT_c/dx = -1.23$ K/$Ce_{\%}$). In its turn, Pr has three *f*-electrons (expected $dT_c/dx \approx -6$ K/$Pr_{\%}$), and Nd has four *f*-electrons ($dT_c/dx \approx -10.5$ K/$Nd_{\%}$ , measured in Ref. [5]). Interestingly, the stabilizing effect of praseodymium should be stronger than that of cerium, so, potentially, adding 15–20% Pr to lanthanum to stabilize and synthesize ternary $(La,Pr)H_{9–10}$ superhydrides might result in obtaining $T_c \approx 100$ K at pressure of 60–65 GPa only. On a similar topic, there is a theoretical prediction on protactinium hydrides ($PaH_8$ and $PaH_9$) [34], where the dynamic and thermodynamic stability of $PaH_8$ ($T_c \sim 79$ K) is found at about 32 GPa. Protactinium is radioactive and unstable, but its analogue among the lanthanides should be praseodymium, albeit with a large number of *f*-electrons, which will undoubtedly worsen the superconducting properties. Thus, praseodymium is a promising additive for reducing the stabilization pressure of lanthanum superhydrides below 65-70 GPa.

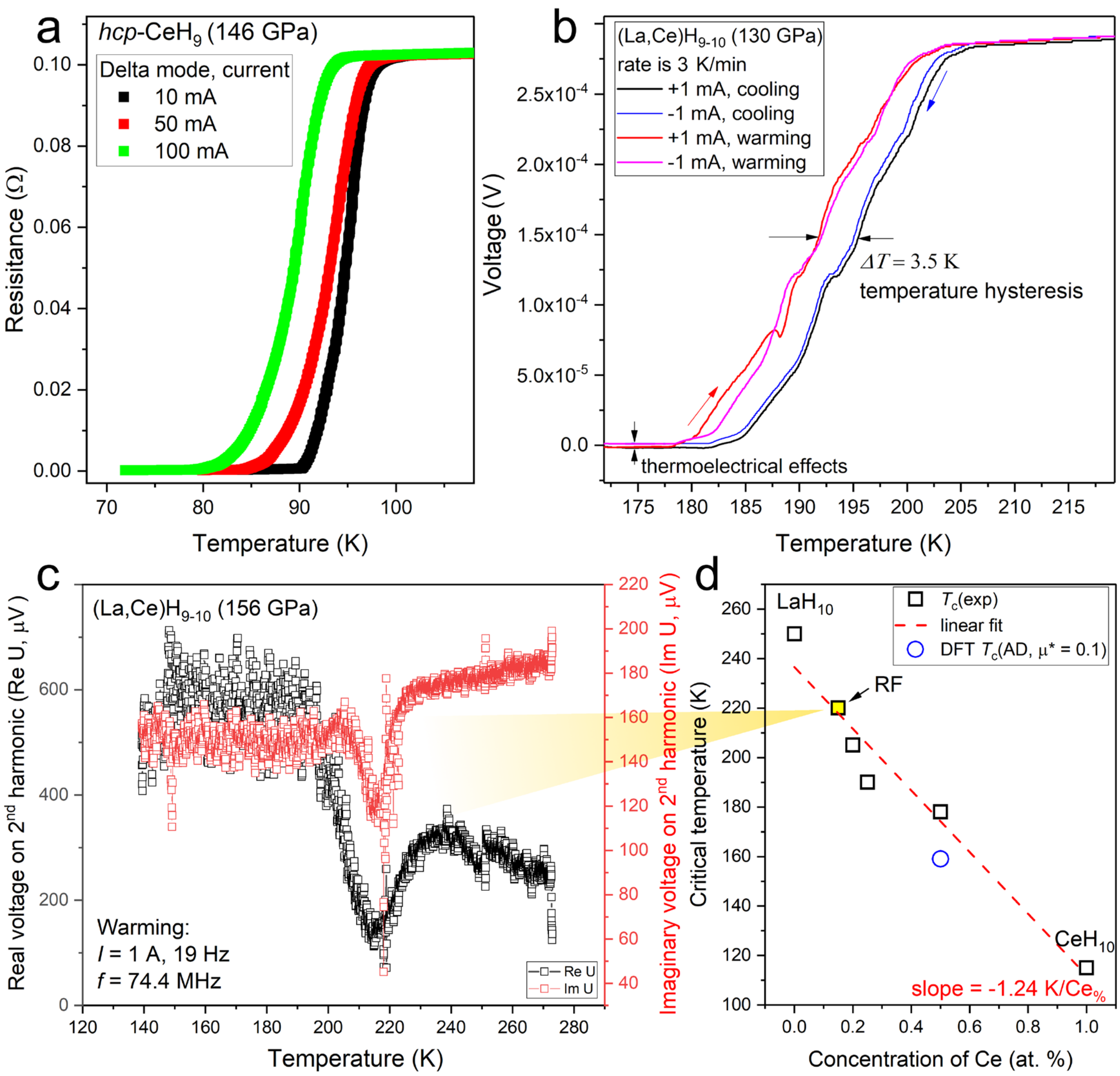

**Figure 3.** Superconducting properties of Cerium and Lanthanum-Cerium polyhydrides. (a) Resistance as a function of temperature for the *hcp*-$CeH_9$ at 146 GPa, measured using three different currents: 10 mA, 50 mA, and 100 mA in the delta mode. The superconducting transition is observed around 95-100 K. (b) Temperature dependence of resistance for $(La,Ce)H_{10}$ at 130 GPa measured in the AC mode ($f$ = 100 Hz) with a DC offset of ± 1 mA, showing superconducting transitions upon cooling and warming. A temperature hysteresis of $\Delta T$ = 3.5 K is indicated, along with the hysteresis due to asymmetric conductivity and influence of thermoelectrical effects. (c) Radio-frequency (RF) transmission measurements on $(La,Ce)H_{10+x}$ sample at 156 GPa in a warming cycle. The real and imaginary parts of the second harmonic signal (solenoid field is 60-70 Gauss, frequency of modulation $F$ = 19 Hz) are plotted against temperature, with a clear feature corresponding to the superconducting transition. (d) Superconducting critical temperature ($T_c$) as a function of Cerium (Ce) concentration for $(La,Ce)H_{10}$. The plot includes experimental data points (black squares) and a theoretical

prediction from Density Functional Theory (DFT) calculations (blue dashed line), showing a near linear decrease in $T_c$ with increasing Ce concentration with a slope of –1.24 K/Ce at.%.

The drawback of the La–Ce alloys is their high sensitivity to oxygen in air. It is almost impossible to avoid significant oxidation of a La–Ce samples in ambient conditions. Thin sputtered films thinner than 1-2 μm oxidize within a few minutes outside a glovebox. A logical development of this system would therefore be to replace part of cerium with another metal of lower chemical reactivity and fewer *f*-electrons, for example thorium [35,36]. Alternatively, one could add to the La–Ce alloy a metal less prone to oxidation in air, such as yttrium. This is the subject of the following section.

### 2.3. Quaternary high-entropy polyhydrides

The development of the La–Ce–H system via additives of yttrium led to the La–Y–Ce–H system (near 1:1:1 composition, max $T_c$ = 180–190 K [37,38]), which can already be classified as a high-entropy hydride due to the significant disorder in the heavy-atom and hydrogen sublattices. Despite the presence of disorder in the crystal structure of this hydride, the superconducting transitions are relatively sharp, within 10–15 K [37].

It is important to note not only the uniquely strong vortex pinning in this hydride and the correspondingly record-high upper critical field $H_{c2}(0) \approx 300$ T [37], but also the fact that the introduction of yttrium practically does not change the superconducting critical temperature of La–Ce–H hydrides. Indeed, Y plays a role of a nonmagnetic impurity, and according to Anderson's theorem, its addition to the La-Ce-H system should not change $T_c$. That is exactly what is observed [37].

Attempts to stabilize yttrium hydrides using cerium have not met with much success at present. In the Y–Ce–H system with 1:1 ratio of Y and Ce, the achieved value of $T_c$ = 138 K [39] is only slightly higher than $T_c(CeH_{10})$ = 115 K. This may indicate that even 50 at. % of the nonmagnetic yttrium impurity keeps system near the scope of Anderson's theorem. This result can be understood by a simple calculation: since the $T_c$ of the Y-H system is in the range from 226 K ($YH_6$ [10]) to 243 K ($YH_9$ [9]), the inclusion of 50 at.% of cerium will lead to a drop in $T_c$ by about $50\times(dT_c/dx) = -50\times1.23$ K/$Ce_{\%} \approx 60$ K, that is, up to 166-183 K in acceptable agreement with the experimental max $T_c$ = 138 K. Likely, optimization of the hydride composition and synthesis pressure-temperature conditions is still possible in this case. So far, only a single $Y_{0.5}Ce_{0.5}$ alloy has been investigated.

Slightly better results were obtained for the Ca–Mg–H system ($T_c$ = 182 K) in Ref. [40], where authors started their synthesis attempts from the intermetallic $Mg_2Ca$ and a 1:1 Ca/Mg alloy. Obtained $T_c$ is not surprising, considering the nonmagnetic nature of Mg atoms: within the framework of $CaH_6$ doping, the critical temperature cannot exceed 210-215 K [41,42] but, of course, can be lower. The maximum critical temperature in this system, 182 K, was detected only at very high pressure of 321 GPa, and the sample contained multiple phases. Around 240 GPa, $T_c$ reached only about 140 K, which is significantly lower than $T_c$ of pure $CaH_6$. These results suggest that Mg does not improve the superconducting properties of $CaH_6$ but rather degrades them. Indeed, pure Mg hydrides ($MgH_{\sim 5}$) show weakly expressed superconductivity ($T_c$ < 70 K around 2 Mbar) despite many theoretical DFT calculations predicting $T_c(MgH_6)$ > 200-240 K [43-45]. The discrepancy between observed and predicted superconductivity in magnesium hydrides is one of the modern mysteries of superhydride physics.

**2.4. Suppression of superconductivity by f-elements: La-Nd-H and Nd-Ca-Zr-H systems**

The suppression of superconductivity by magnesium in the Ca–Mg–H system at high Mg concentrations (50–66 at. %) clearly goes beyond Anderson's theorem and is explained by the fact that magnesium hydrides are poor superconductors, contrary to numerous DFT calculations predicting the opposite [43-45]. Indeed, the critical temperature in disordered ternary hydrides should be a continuous function of the content of a dopant $T_c(x)$. This function should take the values $T_c(x = 0)$ and $T_c(x = 1)$ as in pure binary systems. Moving from the calcium polyhydrides to the magnesium polyhydrides, we expect a decrease in experimental $T_c$.

A different situation arises with *f*-elements, such as neodymium, which suppress superconductivity even at low concentrations in the La–Nd–H system ($dT_c/dx$ = –10.5 K/$Nd_{\%}$) without changing the cubic structure of hydrides in this system [5]. Such a high suppression coefficient explains why pure lanthanide hydrides are not superconductors, except in the cases of La, Ce, Yb, and Lu.

In addition to La–Nd–H, we investigated a compound from the quaternary Ca–Nd–Zr–H system with the starting alloy composition $Ca_{0.5}Nd_{0.4}Zr_{0.1}$. This was one of the first reported cases of quaternary hydride synthesis (see Supp. Info. in Ref. [5]). Despite the complexity of the composition, synthesis using ammonia borane at 169 GPa led to the formation of a polyhydride with a surprisingly simple, nearly ideal cubic (*fcc*) solid-solution type lattice (Figure 4a) and composition $(Ca,Nd,Zr)H_{9+x}$. This high-entropy structure remained stable at least down to 135 GPa (Figure 4b).

Unexpectedly, despite the very high neodymium content, the resulting hydride still exhibited a superconducting transition with a partial resistance drop at $T_c$(onset) = 25–30 K (Figure 4c). Even if we assume $T_c(CaH_9)$ ~ 250 K, then just 30 at. % of Nd ($dT_c/dx$ = –10.5 K/Nd$_{\%}$) would be enough to completely suppress superconductivity in the synthesized quaternary hydride. However, we observe that superconductivity persists even at higher Nd content, indicating that the dependence of $T_c$ on magnetic-atom concentration is not linear.

In a magnetic field, (Ca,Nd,Zr)$H_{9+x}$ behaves as a "hard" superhydrides [46,47] with strong pinning, $|dH_{c2}/dT|$ close to unity, and only minor broadening of the transition in magnetic fields. As with many other metal superhydrides, the phase diagram is close to linear: $\mu_0 H_{c2}(T)$ = 18.5 – 0.84×$T$ (Figure 4d). At present, there is no satisfactory universal explanation for the linear dependence of the upper critical field observed in many hydrides [48,49].

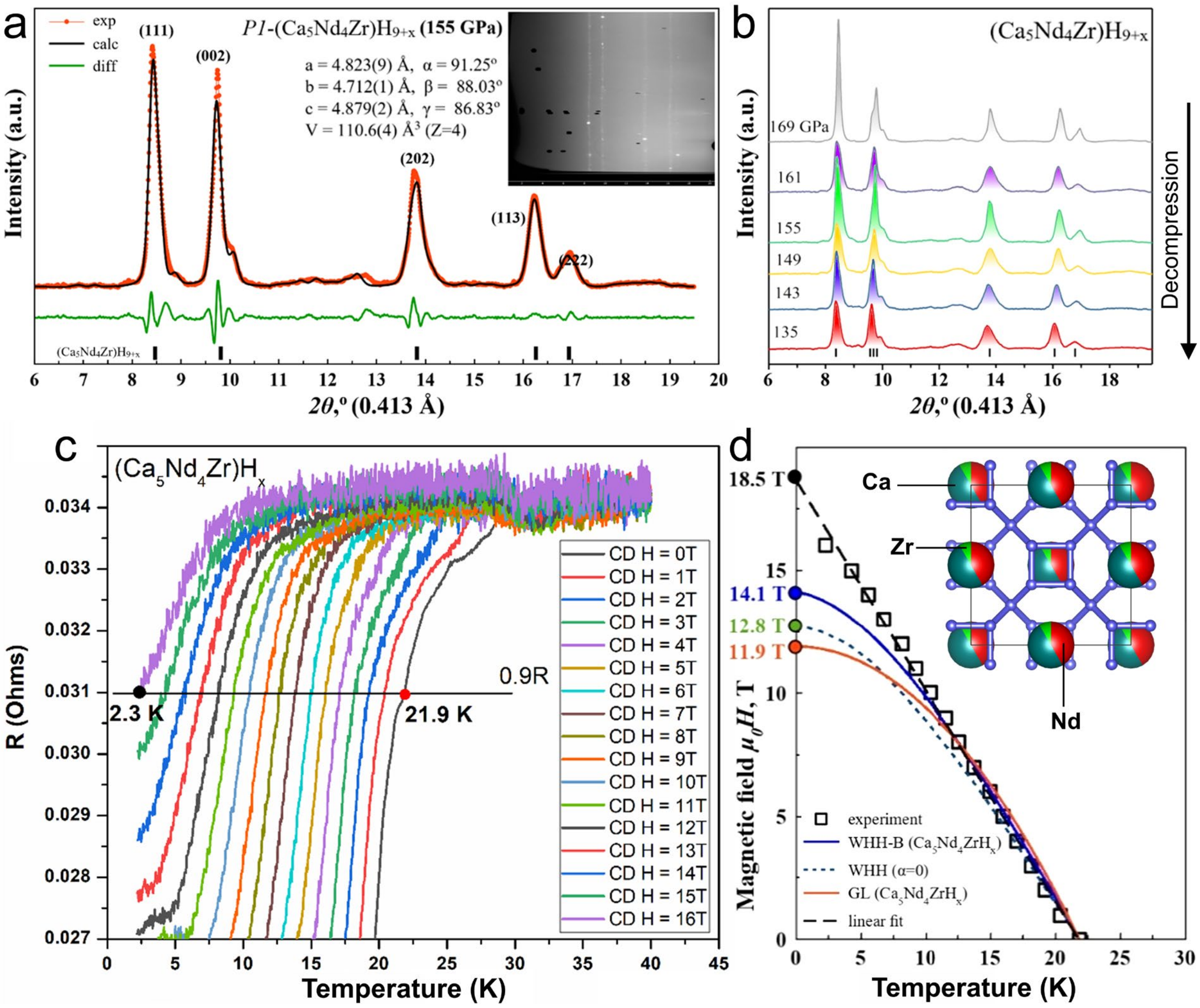

**Figure 4.** Structural and superconducting properties of $Ca_{0.5}Nd_{0.4}Zr_{0.1}H_{9+x}$. (a) Le Bail refinement of the unit cell parameters of of pseudocubic (Ca,Nd,Zr)$H_{9+x}$ at 155 GPa: the X-ray diffraction (XRD) pattern is shown by black line. The red line represents the refinement fit, and the green line is the difference. Despite its closeness to the cubic structure, the crystal lattice is likely distorted due to differences in the atomic radii of the Ca, Nd, Zr atoms. Key diffraction reflections (111, 002, 202, 113, 222) are indexed. The inset shows the raw 2D diffraction image from which the 1D pattern was extracted. (b) A series of XRD patterns of $(Ca,Nd,Zr)H_{9+x}$ obtained during decompression from 169 GPa down to 135 GPa, showing the evolution of the diffraction peaks. (c) Electrical resistance as a function of temperature for $(Ca,Nd,Zr)H_{9+x}$ near the onset of superconducting transition under various applied magnetic fields (0 – 16 T). A $R_{90\%}$ criterion was used to determine $\mu_0 H_{c2}(T)$. (d) Superconducting phase diagram of $(Ca,Nd,Zr)H_{9+x}$, showing the upper critical magnetic field as a function of temperature. Experimental data (black squares) are compared with theoretical fits from the Werthamer-Helfand-Hohenberg (WHH [50],) model and the Ginzburg-Landau (GL [51]) model, along with a linear fit. The inset depicts a possible crystal structure of pseudocubic $(Ca,Nd,Zr)H_{9+x}$, indicating the positions of Ca, Nd, and Zr atoms in the solid solution.

### **2.5.** *Possible room-temperature superconductivity in the La-Sc-H system*

During the study of the La-Sc-H system in 2024-2025, we noticed the presence of an additional step at 274 K in the temperature dependence of electrical resistance *R(T)* of the $(La,Sc)H_x$ sample synthesized from the La-Sc alloy (1:1) and ammonia borane at 189 GPa [28]. Since this DAC was designed for studies in pulsed magnetic fields, only very limited diffraction information was obtained from this sample. XRD indicated the possible presence of a new hexagonal phase *P*6/*mmm*-$(La,Sc)H_{6-7}$ with a unit cell volume of 22.4 Å$^3$/(La,Sc) [28]. Additional contactless radio-frequency transmission measurements [30,52] revealed several anomalies at 242 K, 249 K, and 279 ± 10 K. The first two correspond to superconducting transitions in $(La,Sc)H_{10-12}$ with classical cubic (*fcc*) and possibly hexagonal (*hcp*) structures. According to the Anderson's theorem, such disordered hydrides with solid-solution type of sublattice of heavy atoms, even with increased content of hydrogen (up to 12) due to randomly distributed $H_2$ molecules and H-atoms, cannot have $T_c$ higher than that of the prototype structure, i.e., *fcc* $LaH_{10}$. However, the situation changes when we have another prototype, *P*6/*mmm*-$XH_8$, proposed in the form of $LaSc_2H_{24}$ in the paper of He et al. [53]. To understand what to expect from *P*6/*mmm*-$(La,Sc)H_8$, it is necessary to know the $T_c$ for the similar hexagonal $LaH_8$ and $ScH_8$, however, experimental or theoretical study of these structures has never been done.

Later, Song et. [54] further investigated the La-Sc-H system by delving into the pressure region above 2–2.5 Mbar and found that around 260 GPa, laser heating of La-Sc (1:2) alloy or a mixture of magnetron-sputtered La and Sc nano/micro particles leads to the formation of a thermodynamically stable true ternary superhydride $LaSc_2H_{24}$, which has a larger cell volume, 24.65 Å$^3$/(La,Sc) at 194 GPa (Figure 5a), than our synthesized *P*6/*mmm*-(La,Sc)$H_{6-7}$, and has a pronounced diffraction reflection (100), indicating the presence of two different ordered sublattices of La and Sc. In the case of disordered (La,Sc)$H_{6-7}$, this peak is absent, and the previously suppressed reflection (001) becomes pronounced instead. It is important to note that at 0 GPa, the La-Sc system contains no intermetallic compounds, but at higher pressures, between 50 and 100 GPa, they appear, according to our preliminary USPEX calculations (Figure 5b, c). For this reason, another approach to synthesizing ordered ternary polyhydrides can be proposed for systems lacking intermetallic compounds: first, compress and heat the alloy with a laser, inducing its ordering at high pressure, and then use this alloy in another high-pressure DAC to react with hydrogen (see formula 3, LH – is the laser heating).

$$(La, Sc)\ alloy \xrightarrow{LH\ at\ 50-100\ GPa} LaSc_2 \xrightarrow{LH\ with\ H_2\ at\ 150-250\ GPa} LaSc_2H_{24}\ . \qquad (3)$$

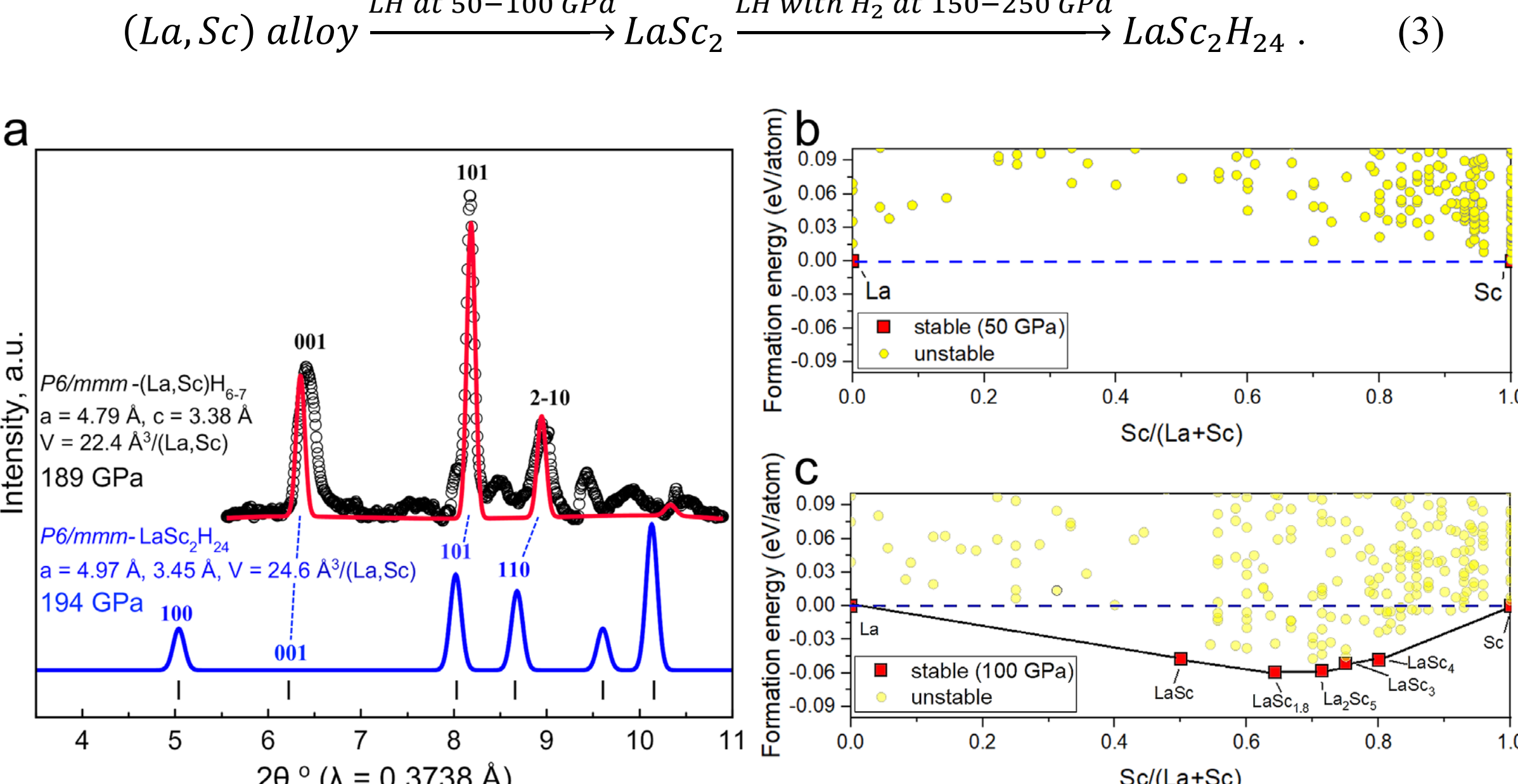

**Figure 5.** Synthesis of ordered hydrides in the La-Sc-H system. (a) Superposition of integrated diffraction patterns and lattice parameters of *P*6/*mmm*-(La,Sc)$H_{6-7}$ and *P*6/*mmm*-$LaSc_2H_{24}$ at 189 and 194 GPa, respectively. It can be seen that, except for small differences in volume and unit cell parameters, both compounds have very similar structures. (b, c) Convex hulls of the La-Sc system constructed at 50 and 100 GPa using USPEX [55] calculations (10 generations of 60 structures, enthalpy was calculated using VASP code[56,57]). It can be seen that applying pressure potentially greatly increases the richness of the La-Sc phase diagram.

## 3. Discussion

The above-mentioned examples of ternary systems with solid-solution sublattices of heavy atoms show that the maximum $T_c$ achieved in $LaH_{10}$ and $YH_9$ is unsurpassed from a practical point of view. From the perspective of raising $T_c$, further efforts in synthesis and research of such disordered ternary systems appear not very promising. A more reasonable goal is increasing the stability of polyhydrides. In particular, encouraging results have been recently predicted for the La–Th–H [58] and Ce–Th–H [59] systems. The two stabilizing atoms, cerium and thorium, provide some hope for obtaining superconducting ternary hydrides at pressures below 65-70 GPa.

Another interesting goal is to increase and experimentally measure the upper critical field in "dirty" high-entropy hydrides (for example, $H_{c2}$ for La–Y–Ce–H can exceed 300 T [37]), and to increase their critical current density for the development of high-current devises based on compressed hydrides.

In Table 1, we present the results of several additional experimental attempts to improve the superconducting properties of superhydrides by adding metals and non-metals into the best binary hydrides. As can be seen, in none of the cases with the possible exception of beryllium, was it possible to achieve interesting practical results.

It should be noted in Table 1 that the experimental work on the synthesis of Y-S hydrides [60] was carried out at 199 GPa, which is very close to the decomposition pressure of $YH_9$. The latter phase can explain high $T_c$(onset) ≈ 236 K, but it could be missed during X-ray analysis due to its decomposition.

**Table 1.** Some recently experimentally studied ternary superhydrides, their synthesis pressures, and their critical temperatures. Critical temperature everywhere is $T_c$(onset). Uncertainty of the $T_c$ determination is ± 5 K

| System | Proposed structure | Pressure, GPa | Experimental $T_c$, K | Reference |
|---|---|---|---|---|
| Y-S-H | *bcc*-(Y,S)$H_{6+x}$ | 199 | 236 K < $T_c$ ($YH_9$) | [60] |
| Y-Ca-H | *bcc*-(Ca,Y)$H_6$ | 148-165 | 224 K < $T_c$ ($YH_6$) | [61] |
| La-Al-H | *hcp*-(La,Al)$H_{10}$ | 164 | 230 K < $T_c$ ($LaH_{10}$) | [62] |
| La-C-H | *fcc* $XH_{10}$ | 172 | 240 K < $T_c$ ($LaH_{10}$) | [63] |
| La-Be-H | *fcc*-$LaBeH_8$ | 81 | 110 K > $T_c$ (*bct*-$LaH_4$) | [64] |
| La-B-H | *I*4/*mmm*-$LaB_2H_8$ | 90 | 106 K > $T_c$ (*bct*-$LaH_4$) | [65] |
| La-Ga-H | *fcc* $XH_{10}$ | 172 | 249 K < $T_c$ ($LaH_{10}$) | [66] |
| La-Ca-H | *fcc* $XH_{10}$ | 173 | 247 K < $T_c$ ($LaH_{10}$) | [67] |
| La-Sc-H | *fcc* $XH_{10}$ | 189-196 | 246 K < $T_c$ ($LaH_{10}$) | [28] |
| La-Sc-H | *P6/mmm*-$LaSc_2H_{24}$ | 189-266 | 274-298 K > $T_c$ ($LaH_{10}$) | [28,54] |

Regarding the La-Be-H system [64], beryllium stabilizes the cubic *fcc*-$XH_4$ structure, which does not occur in the pure binary La-H system, but it appears, for example, in the Sn-H system [49]. Similarly, Anderson's theorem can not be applied to *I*4/*mmm*-$LaB_2H_8$ since it is an ordered truly ternary hydride, where boron occupies specific sites in the crystal lattice [65]. These two La-based compounds containing boron and beryllium offer a possible way out of the impasse associated with disordered ternary systems and Anderson's theorem. This is the synthesis of fully ordered ternary polyhydrides.

Fully ordered ternary hydrides are precisely the class of systems where DFT calculations can demonstrate their full power. Of course, we can expect that in the case of a large difference between the properties of two metal atoms, laser heating to a high temperature will not lead to their mixing in the metal sublattice, but will lead to the formation of a new true ternary polyhydride with fully ordered structure. It is assumed that this is exactly what happened in the case of $LaB_2H_8$ [65]. One must remember however, that heavy atoms with a large difference in their atomic properties from the own atoms in superhydrides can not only replace each other (in solid solution), but also occupy voids in the lattice, and replace hydrogen in some positions, and do this in a disordered manner.

One of the options is prolonged annealing of polyhydrides in an infrared laser beam with a gradual decrease of its intensity. This will not only promote the growth of larger microcrystals of hydrides but also allow thermodynamically stable ternary polyhydrides to form naturally. The currently proposed method for the synthesis of $LaSc_2H_{24}$ belongs precisely to this path. This approach, however, carries the risk of damaging the diamonds during repeated, long-duration, high-intensity laser pulses.

It seems to us that another approach is simpler: "cold" (300-600 K) synthesis of hydrides without disrupting the ordered structure of the initial intermetallic compounds with gradual diffusion of hydrogen into the lattice voids under pressure (Figure 6). This cold synthesis can be hot enough to decompose $NH_3BH_3$ (200-300 °C at ambient pressure), but not hot enough to melt the intermetallic lattice. In this case, the diffuse penetration of hydrogen into the compound can be accelerated by long-term (2-4 weeks) maintenance of the diamond cells at 100-200 °C with periodic measurement of the temperature dependence of the electrical resistance. The feasibility of this process was demonstrated in the works of M. Eremets's group back in 2018, when maintaining DACs with lanthanum and hydrogen for several weeks at room temperature made it possible to obtain superconducting $LaH_{10}$ without using laser heating [6].

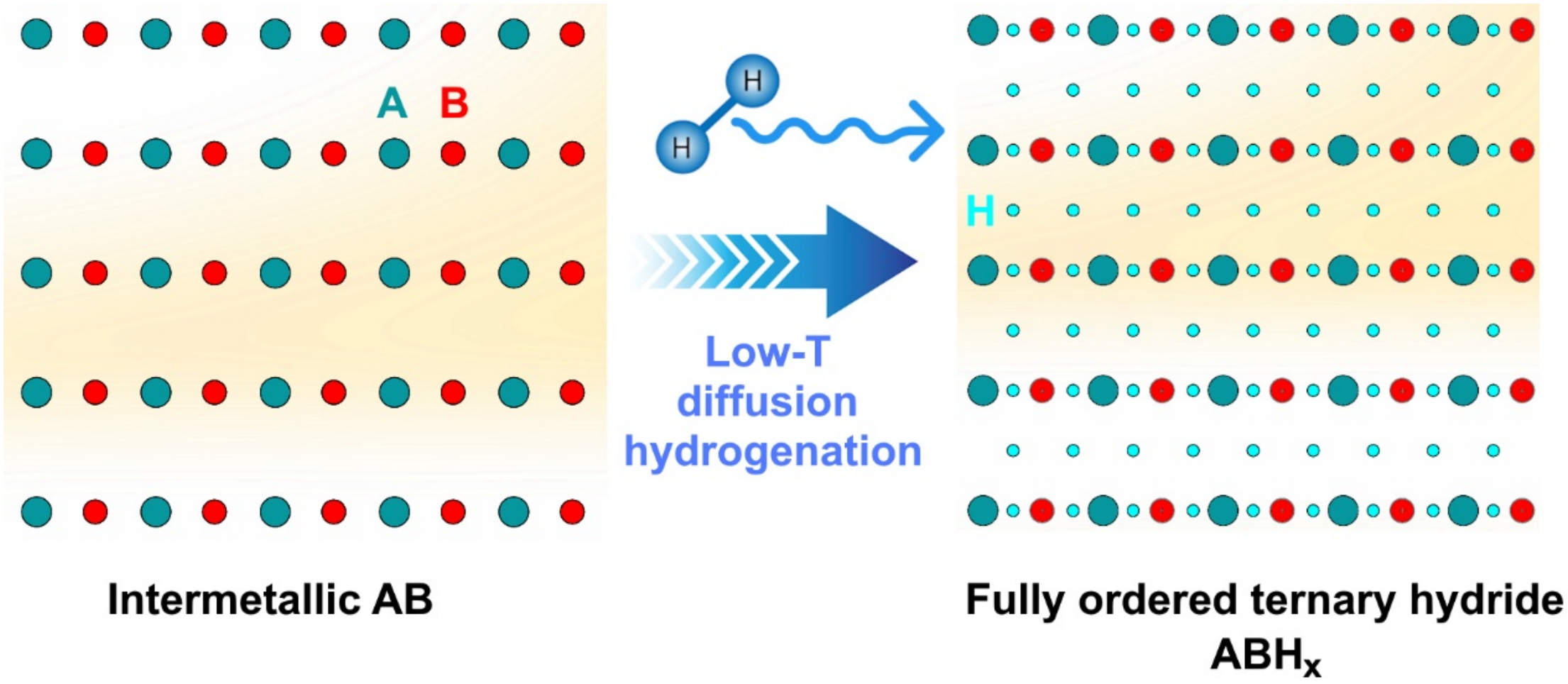

**Figure 6.** Illustration of the "cold" synthesis of superhydrides from completely ordered intermetallics by slow diffusion of hydrogen at high pressure.

As is known, the most potentially interesting ternary hydride systems with the highest $T_c$'s are a simple combination of the "best" binary superconducting hydrides. In other words, combining 12 elements: Mg, Ca, Sr, Ba, La, Y, Sc, Ce, Th, Zr, Hf, Lu we get 66 potentially interesting systems. If we also include alkali metals (Li, Na, K), the number of interesting metallic systems increases. To study these ~ 100 ternary systems even by DFT methods, many years are needed. There are two possible ways that can be considered. One is the development of new computational approaches allowing the fast and accurate simulations of electronic properties. Such methods [68-71], based on machine learning approaches, can be used to effectively perform the electronic structure calculations, and also can be further utilized to solve Eliashberg equations. Another way is to significantly reduce the number of potentially interesting systems. What if we study only the hydrogenation of intermetallics, the number of which is significantly smaller? Some examples are given in Table 2. We believe that the transition to the study of intermetallics will significantly narrow the field of experimental and theoretical research.

**Table 2.** Potentially interesting initial intermetallics and their structures, as well as melting temperatures at ambient pressure

| Formula | Structure | Melting point, °C | Reference |
|---|---|---|---|
| $La_3Ga$ | $Pm\bar{3}m$ | 550 | [72] |
| LaGa | *Cmcm* | 1100 | [72] |
| LaMg | $Pm\bar{3}m$ | 745 | [73] |
| $Y_5Mg_{24}$ | $I\bar{4}3m$ | ≈570 | [74] |
| YMg | $Pm\bar{3}m$ | 927 | [74] |

## 4. Conclusion

Impurities and doping are useful for stabilizing superhydrides at lower pressures, improving vortex pinning, and increasing $J_c$ and $H_{c2}$ in the superconducting state. We believe that it is technically feasible to obtain superhydrides with maximum $T_c$ = 100-150 K at pressures of 60-70 GPa, in other words, below the jamming and cracking limit of diamond anvils. Further attempts to increase $T_c$ by introducing a third element into random sublattice positions of the 'best' binary superhydrides ($LaH_{10}$, $YH_9$, $CaH_6$) appear unlikely to succeed, reminiscent of the insensitivity of conventional superconductivity to nonmagnetic disorder described by Anderson's theorem.

The promising way to increase hydrides' $T_c$ is the "cold" synthesis of ternary polyhydrides based on completely ordered intermetallics. This would close the gap between the very high critical temperatures predicted by density functional theory and the significantly more modest results of the experimental studies of (pseudo)ternary superhydrides with solid solution sublattices.

**Acknowledgements**

D. V. S. and D. Z. thank the National Natural Science Foundation of China (NSFC, grant No. 12350410354) and the Fundamental Research Funds from HPSTAR for support of this research. D. V. S. and D. Z. further acknowledge support under the EMFL-ISABEL project (No. 871106). V.V.S. acknowledges the financial support from Shanghai Science and Technology Committee, China (No. 22JC1410300) and Shanghai Key Laboratory of Materials Frontier Research in Extreme Environments, China (No. 22dz2260800). This work was supported by the National Key Research and Development Program of China (grant 2023YFA1608900, subproject 2023YFA1608903). A.V.S. acknowledges the financial support of RSF grant No. 24-72-10109. We would like to thank staff of the SPring-8 synchrotron, especially Prof. Katsuya Shimizu and Dr. Yuki Nakamoto (Osaka University, Japan) for their assistance in diffraction studies of the Ca-Nd-Zr-H system. We would like to express our gratitude to Dr. Alexey A. Bykov, Dr. Konstantin Y. Terent'ev, Dr. Alexander V. Cherepakhin for the synthesis of Ca-Nd-Zr alloy.

**Data Availability Statement**

The data that support the findings of this study are available from the corresponding author upon reasonable request.

Analysis of experiments with ternary superhydrides reveals the reason why their critical temperatures do not surpass Tc's of the best binary hydrides. The rigidity of superconducting Tc's is a direct consequence of Anderson's theorem for conventional superconductors. While impurities are useful for stabilizing disordered ternary polyhydrides at lower pressures, we argue that further progress requires synthesizing of fully ordered structures.

**Stability and Superconductivity of Ternary Polyhydrides**

*Dmitrii V. Semenok, Di Zhou, Wuhao Chen, Alexander G. Kvashnin, Andrey V. Sadakov, Toni Helm, Pedro N. Ferreira, Christoph Heil, Vladimir M. Pudalov, Ivan A. Troyan, and Viktor V. Struzhkin*

ToC figure

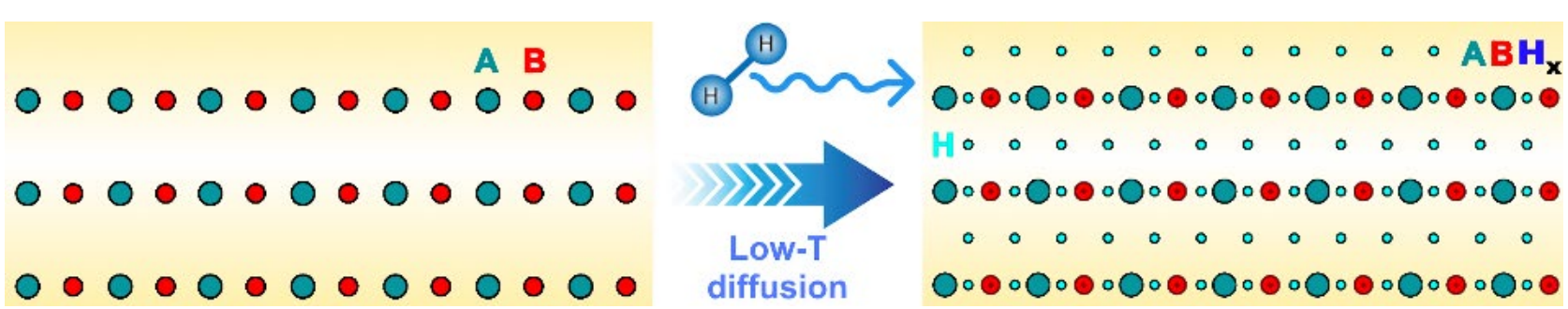